# Energy Dissipation during Diffusion at Metal Surfaces: Disentangling the Role of Phonons vs. Electron-Hole Pairs

Simon P. Rittmeyer,[1,][*] David J. Ward,[2] Patrick Gütlein,[1] John Ellis,[2] William Allison,[2] and Karsten Reuter[1]

[1]*Chair for Theoretical Chemistry and Catalysis Research Center,*
*Technische Universität München, Lichtenbergstr. 4, 85747 Garching, Germany*
[2]*Cavendish Laboratory, University of Cambridge, Madingley Road, Cambridge, CB3 0HE, United Kingdom*

Helium spin echo experiments combined with *ab initio*-based Langevin molecular dynamics simulations are used to quantify the adsorbate-substrate coupling during the thermal diffusion of Na atoms on Cu(111). An analysis of trajectories within the local density friction approximation allows the contribution from electron-hole pair excitations to be separated from the total energy dissipation. Despite the minimal electronic friction coefficient of Na and the relatively small mass mismatch to Cu promoting efficient phononic dissipation, about $(20 \pm 5)\%$ of the total energy loss is attributable to electronic friction. The results suggest a significant role of electronic non-adiabaticity in the rapid thermalization generally relied upon in adiabatic diffusion theories.

Energy dissipation during surface dynamical processes at solid surfaces has been extensively studied, both due to its paramount technological importance and intriguing fundamental richness. Scattering or adsorption of molecules, diffusion and chemical reactions are all known to be intricately governed by the detailed ways in which chemical and kinetic energy is transferred into and out of substrate degrees of freedom. On insulating or semiconducting surfaces the dynamical coupling to the surface can be attributed to the excitation of and interaction with lattice vibrations with some confidence. In contrast, on metal surfaces the role of competing electronic non-adiabatic effects such as electron-hole (*eh*) pair excitations is a continuing topic of debate. In fact, there is growing experimental evidence that can only be rationalized by breaking with the prevalent Born-Oppenheimer view [1, 2]. It may even be argued that due to the continuum of substrate electronic states at the Fermi edge, no dynamical process can strictly be adiabatic at metal surfaces at all [3, 4]. On the other hand, many phenomena still seem to be very well described using purely adiabatic theories [5–9].

Recent *ab initio* calculations of dynamical phenomena beyond the Born-Oppenheimer approximation have attempted to resolve some of this ambiguity [10–14]. In particular the numerically appealing concept of electronic friction [10, 15–17] within the local density friction approximation (LDFA) [18, 19] has become a popular approach in this regard [14, 18, 20–24]. Scattering processes [8, 14, 18, 25, 26] and (dissociative) adsorption events [10, 14, 23, 27] have gained the most attention in this context and with the high incident energies, short contact times and massive charge rearrangements such processes are likely to be good candidates for a high degree of electronic non-adiabaticity.

In comparison to scattering and adsorption processes, the situation is less clear for surface diffusion. On the one hand, diffusing adsorbates are necessarily close to the surface and in regions of high electronic density, with a concomitant amount of electronic friction. On the other hand, the comparably low velocities that are involved may suppress the non-adiabatic channel and thus favor a coupling to the phononic degrees of freedom to finally render surface diffusion electronically adiabatic. Interestingly, a significant contribution of non-adiabatic energy dissipation in the transient H-atom diffusive motion following $H_2$ dissociation over Pd(100) has been reported by Blanco-Rey and co-workers only recently [20, 21, 24]. The results are consistent with a similar prediction by Wahnström made for H diffusion on Ni(100) in the late 1980s [28]. Hydrogen diffusion is, however, a somewhat special case given that competing phononic couplings are small for this very light adsorbate [21, 24].

In order to obtain a more comprehensive insight into the relative importance of lattice vibrations and *eh*-pair excitations for the energy dissipation during surface diffusion we therefore address the thermal motion of Na on Cu(111). Alkali metal systems have long been used as prototypical systems due to the relative simplicity of their surface chemistry [29, 30] and the Na/Cu(111) combination chosen for the current work benefits from having a much higher adsorbate-substrate mass ratio in comparison to H/Pd. Together with the thermally distributed adsorbate velocities, the coupling to phononic degrees of freedom might be expected to be significantly stronger. Simultaneously, the electronic friction coefficient is a material property that exhibits the well known $Z_1$ oscillations as a function of the atomic number [17, 31, 32]. At any embedding density of interest for surface diffusion, the electronic friction is found to be particularly low for light alkali metals. We might, therefore, expect minimal *eh*-pair excitations during the diffusive dynamics of sodium on a free-electron like metal such as copper. As a consequence one would expect phononic coupling to dominate the overall dynamic interaction with the substrate for Na/Cu(111). Analyzing helium spin echo ($^3$He-SE) signatures for surface diffusion with *ab initio*-based Langevin molecular dynamics (MD) simulations we nevertheless find that the energy loss due to electronic friction contributes approximately $(20\pm5)\%$ of the total energy dissipation, thus reinforcing the view that diffusion is an important class of dynamical processes in which electronic non-adiabaticity is anything but negligible.

The helium spin echo technique utilizes the $^3$He nuclear spin as an internal timer, providing direct access to the intermediate scattering function (ISF) $I(\Delta \mathbf{K}, t)$ at a momentum transfer $\Delta \mathbf{K}$ specified by the scattering geometry [33]. As a result of surface adsorbate motion the (auto-)correlation determined through the ISF decays in time, and for processes where the adsorbate couples to the degrees of freedom of the substrate would typically exhibit an exponential decay. The decay rate $\alpha(\Delta \mathbf{K})$ is highly sensitive to the frictional adsorbate-substrate coupling, with a functional dependence on $\Delta \mathbf{K}$ characteristic of the detailed diffusion mechanism [34, 35]. In the present study experiments were conducted at a surface temperature of 155 K with measurements along the $[11\bar{2}]$ azimuth of a Cu(111) crystal dosed to a coverage $\Theta = 0.025$ monolayers (ML) of sodium [36].

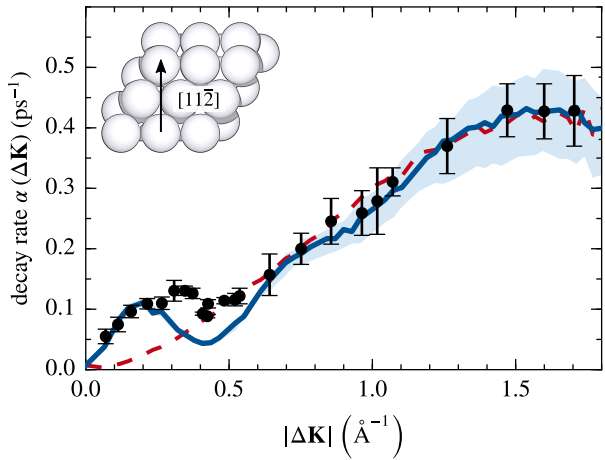

FIG. 1. (Color online) Experimentally measured decay rates $\alpha(\Delta \mathbf{K})$ along the $[11\bar{2}]$ surface direction as opposed to those extracted from simulations with a best-fit friction coefficient of $\eta = 10\,\mathrm{amu\,ps^{-1}}$ (solid blue line). The blue-shaded region indicates the sensitivity when varying the free parameter $\eta$ by $\pm 30\%$. Simulations using an optimum value for $\eta$, but without adsorbate-adsorbate interaction potentials yield the dashed sinusoidal red line. This line lacks the de Gennes narrowing peak at small $|\Delta \mathbf{K}|$, but is unaffected in the region sensitive to the frictional coupling.

The form of $\alpha(\Delta \mathbf{K})$ extracted from the data is shown in Fig. 1. At large values of $|\Delta \mathbf{K}|$ the behavior is indicative of single-jump diffusion, consistent with the Chudley-Elliott model [37] (dashed line), while at smaller values below about $0.6\,\mathrm{\mathring{A}}^{-1}$ there is an obvious deviation from the ideal sinusoidal signature that is consistent with "de Gennes narrowing" [38] and observed for previous works on repulsive interacting adsorbates [35], notably sodium diffusing on the Cu(100) surface [39].

A quantification of the adsorbate-substrate frictional coupling can be achieved within the kinematic scattering approximation [35]. As further detailed in the SI [40], the ISF is directly related to the real-space motion $\mathbf{R}_j(t)$ of an ensemble of $N_{\mathrm{atoms}}$ adsorbates $j$ through the autocorrelation function of the coherent intermediate amplitudes

$$A(\Delta \mathbf{K}, t) = \sum_{j}^{N_{\mathrm{atoms}}} \exp\left[-i \Delta \mathbf{K} \cdot \mathbf{R}_j(t)\right] \quad . \quad (1)$$

Corresponding trajectories $\mathbf{R}_j(t)$ are conveniently obtained from Langevin MD simulations, in which the overall friction coefficient $\eta$ is varied until optimum agreement with the experimental decay rates is obtained [34, 35, 41]. Specifically, in the current work we employed a system of $N_{\mathrm{atoms}} = 200$ adatoms in a supercell consisting of an $(49 \times 82)$ array of rectangular Cu(111) unit cells and used $T = 155$ K to match the experimental Na coverage and temperature. Appropriate averaging over 100 MD runs accumulated over 1.6 ns ($2^{14}$ steps) each ensured converged decay rates $\alpha(\Delta \mathbf{K})$.

To minimize the number of free parameters the two-dimensional adsorbate-substrate potential energy surface (PES) employed in the Langevin MD simulations was determined by density-functional theory (DFT) calculations using CASTEP [42] at the generalized gradient PBE functional level [43]. As detailed in the SI [40] these calculations are used to parametrize an analytical Fourier representation of the PES, which faithfully reproduces the DFT PES with a root mean square deviation of $< 2\,\mathrm{meV}$. As indicated by the de Gennes narrowing feature at small $|\Delta \mathbf{K}|$ in Fig. 1, we additionally account for repulsive adsorbate-adsorbate interactions through pairwise repulsive dipole-dipole interaction potentials according to Kohn and Lau [44]. The required (coverage-dependent) dipole moments of the respective adatoms are obtained by fitting experimental work function-change measurements [45] to the Topping model of surface depolarization [46], as had already been done successfully for Na on Cu(100) [47].

The resulting analysis exhibits only one remaining free parameter, the friction coefficient $\eta$. As shown in Fig. 1 an optimized value of $\eta = 10\,\mathrm{amu\,ps^{-1}}$ achieves an overall excellent agreement with the experimental measurements. All prominent features in the experimental curve, i.e. the modulation corresponding to the de Gennes narrowing at small values of $|\Delta \mathbf{K}|$, as well as the sinusoidal line shape for larger values are qualitatively reproduced with the major contributory factors to diffusion quantitatively reproduced to a large extent. To obtain an estimate of the sensitivity of our results, we additionally indicate in Fig. 1 the range of $\alpha(\Delta \mathbf{K})$ values we obtain when varying the best-fit friction coefficient within $\pm 30\%$. It is obviously only the region at $|\Delta \mathbf{K}| > 0.7\,\mathrm{\mathring{A}}^{-1}$ that is increasingly sensitive to this friction coefficient, and the $\pm 30\%$ uncertainty safely brackets the experimental error bars. The small but apparently systematic deviations in the lower $|\Delta \mathbf{K}|$ region are instead attributed to a conceivably insufficient treatment of adsorbate-adsorbate interactions. When completely switching off the dipole interactions in our simulations, the changes to the sinusoidal shape predicted by the single jump-model [37] are exclusively restricted to this low $|\Delta \mathbf{K}|$ region, cf. Fig. 1. Thus, the friction value we obtain is completely robust with respect to these aspects of our model. A similar robustness is obtained with respect to the PES topology. As detailed in the SI [40], variations of the diffusion barrier over the bridge sites by ±30%, to account for inaccuracies of the DFT PBE functional we use, also leads to a variation of decay rates that fall almost exactly within the shaded region in Fig. 1.

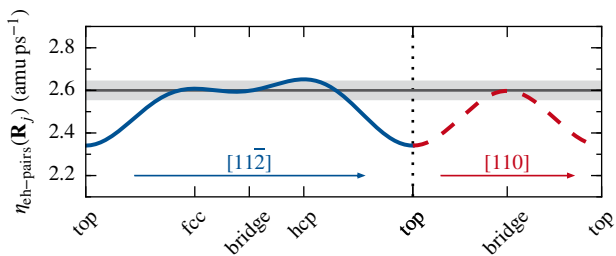

FIG. 2. (Color online) Interpolated electronic friction coefficient $\eta_{eh-\text{pairs}}(\mathbf{R}_j)$ experienced by a Na atom along the $[11\bar{2}]$ (solid blue) and $[1\bar{1}0]$ (dashed red) surface direction. The horizontal dark gray line indicates the determined ensemble- and time-averaged electronic friction $\eta_{eh-\text{pairs}}$, with the light-gray corridor indicating the standard deviation over all time steps and trajectories.

The friction coefficient has contributions from both phononic and electronically non-adiabatic dissipation [48]. In a two-bath model for diffusion, contributions have been shown to be additive [49] so we can write

$$\eta \approx \eta_{\text{phonons}} + \eta_{eh-\text{pairs}} \ . \qquad (2)$$

To disentangle the two dissipation channels approximately, we calculate the ensemble-averaged electronic friction experienced over the Langevin-MD trajectories within the LDFA [15, 17–19]. For this we first determine an analytic Fourier representation of the position-dependent electronic friction coefficient of a diffusing Na atom $\eta_{eh-\text{pairs}}(\mathbf{R}_j)$ using a procedure analogous to that employed for the PES. At each DFT point $\mathbf{R}_{\text{DFT}}$ calculated for the PES parametrization, the embedding density required in the LDFA ansatz is extracted from the self-consistent total electronic density through an atoms-in-molecules scheme based on Hirshfeld decomposition [22]. The resulting grid of $\eta_{eh-\text{pairs}}(\mathbf{R}_{\text{DFT}})$ is subsequently expanded in a Fourier series as further detailed in the SI [40]. Figure 2 illustrates the resulting continuous electronic friction coefficient along two high symmetry lines along the Cu(111) surface. Obviously, $\eta_{eh-\text{pairs}}(\mathbf{R}_j)$ correlates with the inverse height-profile of the Na adsorbate; closer the adsorbate is to the Cu(111) surface, the higher the embedding density and the larger the friction coefficient becomes.

The average electronic friction experienced by the entire Langevin ensemble of adatoms $j$ is then approximated non-selfconsistently at each MD time step as $\eta_{eh-\text{pairs,av}}(t) = \sum_j^{N_{\text{atoms}}} \eta_{eh-\text{pairs}}\left(\mathbf{R}_j(t)\right) / N_{\text{atoms}}$ for each trajectory generated in our best-fit simulations. Averaging over all trajectories and time steps we finally arrive at an estimate of the electronically non-adiabatic dissipation contribution to the overall $\eta$ of $\eta_{eh-\text{pairs}} = 2.60$ amu ps$^{-1}$. As apparent from Fig. 2 this average value is somewhere between the friction coefficients experienced at the most stable fcc and hcp adsorption sites and the lowest-energy diffusion barrier over the bridge sites. As also shown in the figure, the standard deviation resulting from this average $\eta_{eh-\text{pairs}}$ is very small (±0.04 amu ps$^{-1}$), consistent with the fact that the thermalized Na atoms spend the predominant time in the corresponding (meta)stable basins of the PES. In terms of the motion through the surface electron density, the situation is thus highly comparable to vibrational dynamics, an area where the LDFA has been shown to perform quantitatively [22]. Correspondingly, we expect this level of theory to provide a reliable assessment of the relative amount of electronic friction, even though it would be conceptually interesting to compare to higher-level theories that for instance account for tensorial aspects of friction [50] or that additionally provide the explicit $eh$-pair excitation spectra [12, 51]. We further note that similar to the findings for adsorbate vibrations [22], a key element in the use of the simple LDFA scheme is the appropriate determination of the host embedding density experienced by the adsorbate. For the analysis so far, we used the atoms-in-molecules approach based on Hirshfeld's projection scheme [22]. The corresponding integrated Hirshfeld charges indicate a charge transfer of $0.3e$ from a Na atom adsorbed in the fcc or hcp sites to the Cu substrate, which naturally enhances the embedding density and thus the electronic friction coefficient. Use of the independent-atom-approximation as originally employed within the LDFA context [18] does not account for such charge transfer in constructing the embedding density but relies on the self-consistent screening of the underlying isotropic model system. This would then predict an $\eta_{eh-\text{pairs}}$ that is just about 63% of the value determined here. Due to the ambiguous choice of the embedding density, both methods can be considered to yield an upper and lower limit of the LDFA approach, respectively [22].

Given these considerations and comparing the determined $\eta_{eh-\text{pairs}}$ with the total friction coefficient, we arrive at the surprising result that electronic non-adiabaticity amounts to about $(20 \pm 5)\%$ of the total energy dissipation, and this in a system that was selectively chosen to minimize this dissipation channel. Tentatively, we would thus expect even more pronounced influences of $eh$-pair excitations in the diffusive motion of adsorbates like potassium atoms, i.e., elements that correspond to a maximum of the $Z_1$ oscillations of the electronic friction coefficient. As had been shown in the previous work on H diffusion [20, 21, 24, 28], the relative contribution will, of course, also be increased at smaller adsorbate-substrate mass ratios by the concomitant suppression of phononic dissipation. All in all, the picture that emerges is of surface diffusion in which electronic non-adiabaticity plays a much more prominent role than hitherto anticipated. Indeed, one could conjecture that it is in fact electronic non-adiabaticity that ensures rapid thermalization in adsorbate systems with large frequency mismatch and that explains the long-term success of adiabatic theories to determine diffusion constants and other kinetic parameters for growth and catalysis applications.

We thank J.I. Juaristi for providing us with an interpola-tion function for the LDFA electronic friction coefficient of sodium. SPR acknowledges support of the Technische Universität München - Institute for Advanced Study, funded by the German Excellence Initiative.

---

# Energy Dissipation during Diffusion at Metal Surfaces: Disentangling the Role of Phonons vs. Electron-Hole Pairs

Simon P. Rittmeyer,[1, *] David J. Ward,[2] Patrick Gütlein,[1] John Ellis,[2] William Allison,[2] and Karsten Reuter[1]

[1]*Chair for Theoretical Chemistry and Catalysis Research Center,
Technische Universität München, Lichtenbergstr. 4, 85747 Garching, Germany*

[2]*Cavendish Laboratory, University of Cambridge, Madingley Road, Cambridge, CB3 0HE, United Kingdom*

## S1. EXPERIMENTAL DETAILS

### A. Sample Preparation and Characterization

A mechanically polished single crystal Cu(111) sample (Surface Prep. Lab., NL) used in the study is mounted on a sample manipulator, allowing translational, polar and azimuthal rotations as well as temperature control. The manipulator is fitted into a scattering chamber for the spectrometer which is evacuated to $2 \times 10^{-11}$ mbar base pressure post baking. Preparation of the surface consists of repeated cycles of argon ion sputtering ($I_{\text{emiss}} \approx 6\,\mu\text{A/cm}^2$, 800 V Ar$^+$ ions, $T_\text{s} = 300$ K for 30 mins) followed by surface annealing ($T_\text{s} = 800$ K, 30 secs). The surface quality is monitored through measurement of helium reflectivity. A high quality surface was confirmed regularly by exceptionally strong helium reflectivity ($> 34\%$ measured at $T_\text{s} = 300$ K). A typical incident energy of 8 meV was used for the experiments with the beam energy recorded at regular intervals.

The clean Cu(111) crystal was aligned to the $[11\bar{2}]$ surface azimuth, by optimizing the pattern of helium scattered from high purity carbon monoxide adsorbed to monolayer (ML) saturation. The temperature of the sample is monitored using a type-K [1] thermocouple spot welded onto a sample mount constructed from tantalum. Temperature control is achieved with cryogenic sample cooling using liquid nitrogen ($T_\text{s} > 120$ K), balanced against radiative heating from a coiled tungsten filament.

Alkali metals are dosed onto the Cu(111) sample from dispensers supplied by SAES Getters[2], which provide a convenient method for introducing high purity films in vacuum. In order to deliver alkali vapor efficiently, the front edge of the dispenser and the surface must be brought into close proximity. For the current work an apparatus has been constructed consisting of a linear vacuum below with a dosing insert onto which the dispenser is fixed with a titanium flag in front connected to an external rotary vacuum feed-through. When the flag is closed the sample is shielded from the dispenser. Opening and closing the flag allows a precise initiation and termination of dosing irrespective of the dispenser pre-loading conditions. Before starting to dose with a new dispenser it is degassed to remove adsorbed gases from the casing and support mountings.

---

* Corresponding author: simon.rittmeyer@tum.de

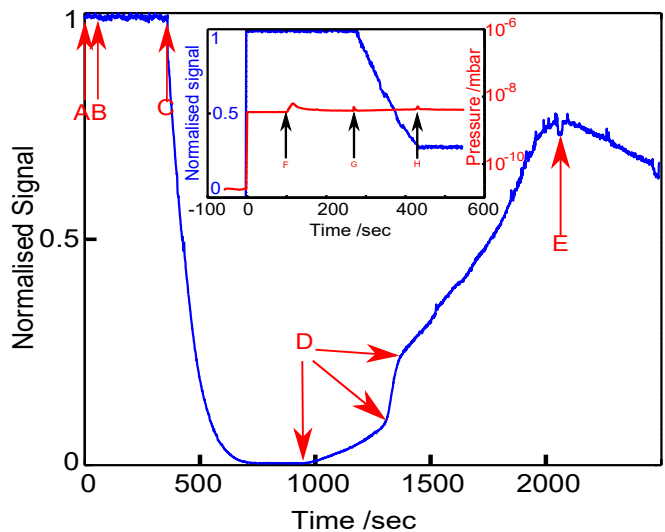

FIG. S1. Na on Cu(111) dosing curve. The gate valve between the scattering chamber and dosing arm is opened at "A" ($t = 0$ s), the dispenser current is enabled at "B" ($t = 50$ s), but no loss of specular reflectivity is observed until the flag is opened at "C" ($t = 350$ s). In the period from "C" onwards sodium is deposited on the surface, demonstrated by an initial decrease in helium reflectivity as the surface entropy increases and then an increase as the surface stabilizes, "D", approaching the complete monolayer, highlighted at "E". At the points indicated by "D" different compressed surface structures are formed. The inset shows helium reflectivity and chamber pressure, in blue and red respectively. In this instance the dose is stopped by closing the dosing flag at "H", where the specular signal is $I_0/3$. The points "F" and "G" indicate the times when the dispensing current is enabled and the flag opened, respectively

Figure S1 shows an uptake curve taken to a coverage greater than monolayer saturation. The period marked "A" through "C" demonstrates that there is no change in specular reflectivity between opening the dosing arm chamber and opening the flag. Sodium is deposited from a clean surface at "C" to monolayer saturation at "E" and beyond. In order to work at a specific coverage the dosage may be stopped virtually instantaneously by closing the flag, as demonstrated in the inset of Fig. S1, where the sample is dosed to a specular attenuation of $I_0/3$.

From the uptake curve (see Fig. S1) and assuming a unity sticking co-efficient, as in the coverage dependent LEED and TPD studies conducted by Tang *et al.* [3], and photo emission spectra in Ref. 4, the coverage can be linearly interpolated from the region "C" through "E" on Fig. S1. The dynamics mea-surements were collected at a coverage corresponding to an attenuation such that $I = I_0/3$, as shown in the inset of Fig. S1, which translates to a coverage of 0.045% of the atoms at saturation coverage. The periodicity of the structure at monolayer coverage is $(3 \times 3)$ but with 4 adatoms per unit cell [3, 5]. So the saturation coverage, defined with respect to the number of substrate atoms in the top most layer is 4/9. Using the known monolayer structure and uptake curve the coverage with respect to the number of substrate atoms in the current study is thus $\Theta = 0.025$ ML.

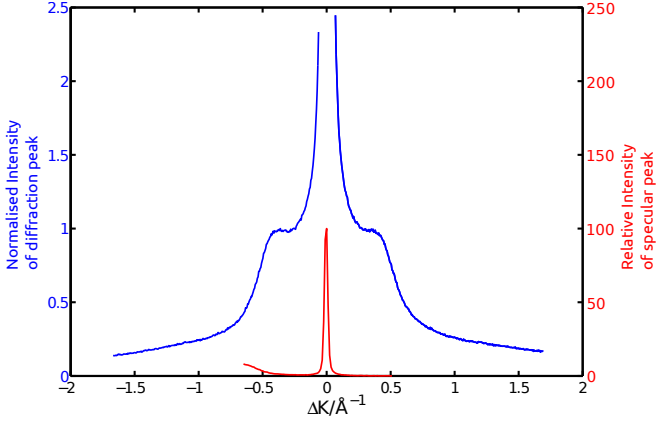

FIG. S2. $^3$He diffraction scan for Na on Cu(111), along the [११$\bar{2}$] substrate direction, at coverage of $\Theta = 0.025$ ML. The signal is normalized to one at the peak of the diffraction ring. The total scattering angle and beam energy are fixed at 44.4° and 8 meV respectively, while the angle of incidence is varied in order to obtain the diffraction pattern. The central peak, shown as a red line, is the specular reflected beam and the weak diffraction features are apparent in the expanded curve (blue line). The diffraction features do not vary significantly with azimuthal sample orientation, indicating an absence of azimuthal ordering.

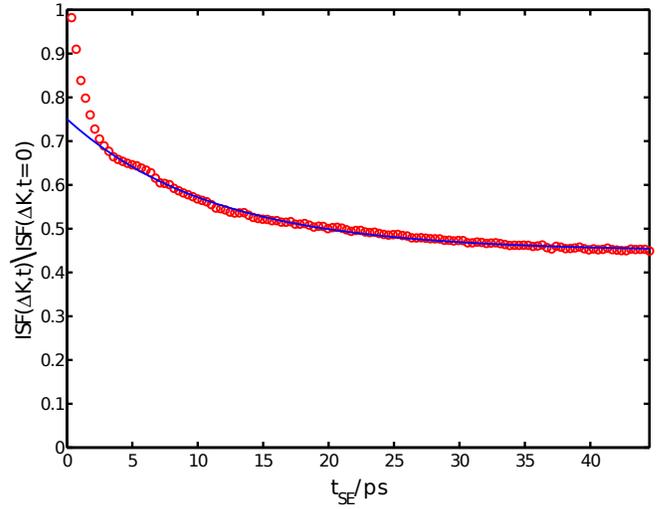

FIG. S3. A typical Na/Cu(111) ISF measured on the [११$\bar{2}$] surface direction and presented for a momentum transfer $\Delta K = 0.08$ Å$^{-1}$ at a coverage of $\Theta = 0.025$ ML and temperature 155 K. The blue line shown is of the form $A_1 \exp(-\alpha t) + A_2$ where $A_1$, $A_2$ and $\alpha$ are determined using a nonlinear least squares fit. The line is in excellent agreement with the data for $t_{SE} > 5$ ps.

The coverage calibration can be cross-checked using the location of diffraction features. An angular intensity scan taken at the same coverage is presented in Fig. S2, which shows a strong, sharp specular signal at $\Delta K = 0$ Å$^{-1}$ together with broader, weaker diffraction-peaks. The observed features correspond to isotropic diffraction rings that result from the quasi-hexagonal distribution of sodium atoms with well defined nearest-neighbor distance but no long-range orientation order. The radius of the inner ring, $K_{ring}$, is related to the average nearest-neighbour distance $r$ by $K_{ring} = 4\pi/\sqrt{3}r$ [6]. The data gives $\Theta = r^2/a^2 = 0.025$ ML which is in excellent agreement with the coverage calculated using the uptake curve.

### B. Measurement and Analysis of the ISF

During measurements care is taken to avoid contamination. The variation of helium-3 reflectivity of the clean copper surface over a period over 5 hours; longer than the maximum measurement session of 3 hours shows no significant variation. ISFs are measured non-sequentially in momentum transfer or temperature, and spectra recorded at the the beginning of each measurement session are repeated at the end with no variation noted.

Figure S3 shows a typical ISF with experimental data shown as red circles. The blue line represents a model of the form $A_1 \exp(-\alpha t) + A_2$, with the free parameters $A_1$, $A_2$ and $\alpha$, optimized using a nonlinear least squares method implemented using the Matlab$^{TM}$ curve fitting toolbox. The model does not represent the data at small times, $t_{SE} < 10$ ps, which are therefore excluded, using an iterative routine to find the optimum exclusion limit. We quantify the quality of the fit using the adjusted coefficient of determination $R^2_{adj}$. If we define the data as a series $y_{i=1}^n$ and the fit as $f_{i=1}^n$, then $R^2_{adj}$ is defined as

$$R^2_{adj} = 1 - \frac{(n-1)\sum\limits_{i=1}^{n}(y_i - f_i)^2}{(n-m)\sum\limits_{i=1}^{n}(y_i - \bar{y})^2}, \quad (S1)$$

where $m$ is the number of degrees of freedom in the model and $\bar{y}$ is the arithmetic mean of the data.

Figure S4 shows the quality of the fit measured using $R^2_{adj}$, and the decay rate $\alpha$ for the blue line shown in figure S3, as a function of the cut off time, starting with the slowest 5 data points and incrementing towards $t = 0$ ps. At large cut-off times, the quality of the fit is limited by the lack of data. Therefore the $R^2_{adj}$ value starts small, and the value of $\alpha$ varies over a relatively large range. As the number of data points increases, $\alpha$ stabilizes around 0.05 ps and $R^2_{adj}$ increases, indicating that the model represents the data. At times less than 12 ps there is a slight reduction in $R^2_{adj}$, which is attributable to the incomplete removal of the inelastic scattering signal in this case. At times less then 5 ps $R^2_{adj}$ decreases rapidly as the quantity of data not described by the exponential model increases, which is consistent with the clear fast decay process in figure S3. A cut off limit for the dataset is found to be 18 ps, yielding a value for $\alpha$ of $0.055 \pm 0.005$ ps$^{-1}$. The same method is applied to analyse the diffusion signal in the rest of the dataset.

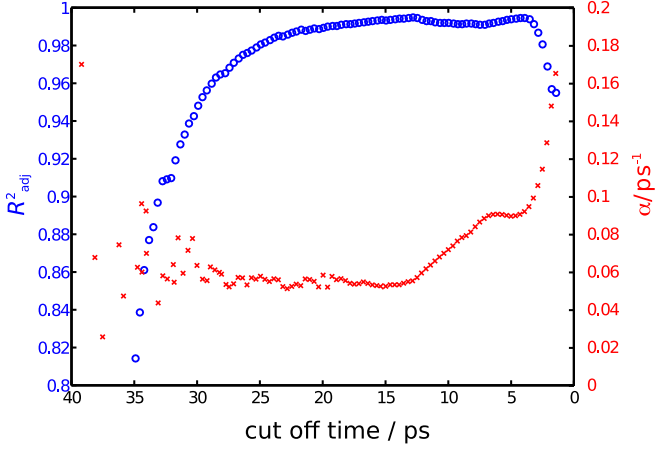

FIG. S4. Sensitivity of the time cut off, when optimizing an exponential model $A_1 \exp(-\alpha t)$ to experimental data (see Fig. S3). Blue circles show the adjusted coefficient of determination $R^2_{adj}$, and red crosses the value of the obtained decay rate $\alpha$ as a function of the cut-off time. It is clear that at large cut off time the fit to the exponential is weak, and as at small times the influence of the clearly defined faster decay distorts the result. The starting point of the current work is 18 ps, yielding a value for $\alpha$ of $0.055 \pm 0.005$ ps$^{-1}$.

The features presented and described in figure S3 are typical for the whole dataset, and can be summarised as a rapidly decaying contribution at times less than 5 ps, followed by a slower decay process. There is a strong inelastic component present in some spectra which is not treated in the current work and does not affect the results presented. It is clear that there is a significant deviation from the exponential fit at small times, typically below 5 ps, which is considered in future work.

## S2. COMPUTATIONAL DETAILS

### A. DFT Calculations and Interpolation of the Interaction Potential

We generate a two-dimensional adsorbate-substrate interaction potential using density-functional theory (DFT) within the generalized gradient approximation (GGA) in terms of the PBE functional [7] as implemented in the plane-wave pseudopotential code CASTEP [8]. In detail, we model the Na on Cu(111) system in a $(3 \times 3)$ surface unit cell, where the metal substrate is represented by 5-layer slabs separated by a 20 Å vacuum layer in $z$-direction between the periodic images. We further employ a plane wave cut-off energy of 400 eV, ultrasoft pseudopotentials [9] and an $(8 \times 8 \times 1)$ Monkhorst-Pack [10] **k**-point grid. For all calculations presented we rely on the frozen surface approximation: We evaluate the optimal clean-surface configuration only once (where the lowermost two layers are constrained to the truncated bulk positions) and subsequently keep all substrate degrees of freedom fixed. To calculate the interaction energies, we then place the Na atom at a defined $(x, y)$ position in the surface unit cell and fully relax its $z$-coordinate using the BFGS algorithm with a residual force tolerance of $0.05$ eV/Å. All computational parameters have been carefully tested to yield interaction energies that are converged to within $< 5$ meV.

To obtain a continuous analytical description of energies and forces, we use the calculated Na interaction energies at the four high-symmetry sites of the Cu(111) surface (top, bridge, fcc and hcp) and expand these in a truncated Fourier series

$$V(\mathbf{R}) = \sum_{i,n} A_n \cos\left(n\mathbf{g}_i \cdot \mathbf{R}\right) + \sum_{i,m} B_m \sin\left(m\mathbf{g}_i \cdot \mathbf{R}\right) + C \quad , \tag{S2}$$

Here, we use two cosine, one sine component (that in principle allows to distinguish between non-degenerate hcp and fcc hollow sites) and a constant offset, such that all parameters are uniquely determined by the four input energies shown in Tab. S1. We further use a redundant set of reciprocal lattice vectors

$$\mathbf{g}_i = \frac{4\pi}{\sqrt{3}a_{Cu(111)}} \begin{pmatrix} \cos(\varphi_i) \\ \sin(\varphi_i) \end{pmatrix}; \quad \varphi_i \in \left[0, \frac{\phi}{3}, \frac{2\pi}{3}\right] \quad, \tag{S3}$$

where the optimized Cu(111) surface lattice constant is $a_{Cu(111)} = 2.55$ Å within our computational setup. Our so-gained analytic interaction potential exhibits an RMSD value of $< 2$ meV as compared to a test set of 231 explicitly calculated DFT relative interaction energies at $(x, y)$ Na-positions covering the entire irreducible wedge of the primitive surface unit cell (see Fig. S5).

### B. Pairwise Interaction Potential

On top of the *ab initio*-based adsorbate-substrate interaction potential we add pairwise repulsive dipole-dipole interactions in our molecular dynamics (MD) simulations. The same pair potentials have been shown to yield convincing results in previous studies of Na on Cu(100) [11]. Other interactions, for example those mediated by surface-states [12], are oscillatory in nature and are significantly weaker so we do not include them here. According to Kohn and Lau [13], the interaction potential of two dipoles with dipole moments $\mu_i$ and $\mu_j$ on a metal surface is given by

$$V^{\mathrm{KL}}_{ij}(\mathbf{r}) = 2\mu_i\mu_j \frac{\mathbf{r}}{r^4}, \tag{S4}$$

where $\mathbf{r}$ is the distance vector between the dipoles. The additional factor of 2 as compared to the classical dipolar interaction energy in vacuum accounts for image charge effects.

TABLE S1. Relative DFT interaction energies of Na on Cu(111). The hcp and fcc hollow sites are energetically degenerate. These four energies are the input values for the Fourier expansion of the analytic interaction potential in Eq. (S2).

| site | relative interaction energy (meV) |
|---|---|
| top | 96 |
| bridge | 12 |
| hcp | 0 |
| fcc | 0 |

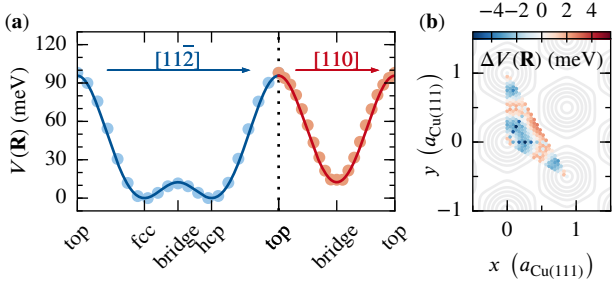

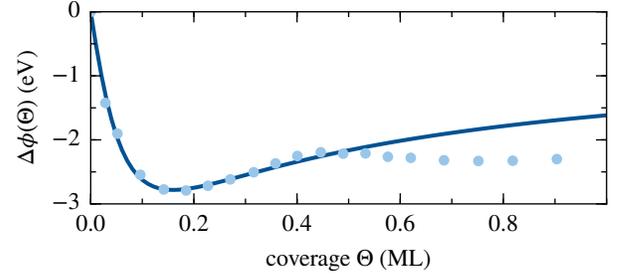

FIG. S5. **(a)** Interpolated Na-Cu(111) interaction potential along two high symmetry lines along the Cu(111) surface (drawn lines) as compared to respective DFT data points from the test set. **(b)** Deviations $\Delta V(\mathbf{R}) = V(\mathbf{R}) - V_{\mathrm{DFT}}(\mathbf{R})$ of the interpolation function from the entire test set of 231 data points in the irreducible wedge of the surface unit cell. The top position is located at $(0,0)$. We obtain an RMSD value of 1.9 meV for the entire test set.

FIG. S6. Fit (dark drawn line) of the work function change with coverage within the Topping model (cf. Eq. (S5)) to experimental measurements (light circles) by Fischer et al. [16]. Only values $\Theta < 0.25$ ML enter the fitting procedure. Note that the coverage definition in this work differs from that in the original work in Ref. 16.

The respective dipole moment of the adsorbed Na atoms is coverage-dependent; the closer the packing is the smaller the dipole moments. An analytical model to include the underlying dipole-induced surface depolarization effects has been proposed by Topping [14]. Treating the adsorbate layer within a plate capacitor-model, the work function change of the substrate as induced by the adsorbates is given by [15],

$$\Delta\phi(\Theta) = -\frac{n_0 \Theta \mu_0}{\varepsilon_0 \left[1 + 9\alpha\left(n_0\Theta\right)^{3/2}\right]}, \tag{S5}$$

where $\alpha$ is the adsorbate polarizability and $\mu_0$ the adsorbate dipole moment in the zero-coverage limit. The adsorbate density per unit surface area at full coverage $\Theta = 1$ is $n_0$. We define the coverage $\Theta$ as number of adsorbates per surface substrate atom. Hence, for a hexagonal (111) surface $n_0 = 2/\sqrt{3}a^2$, where $a$ is the surface lattice constant. The remaining free parameters $\alpha$ and $\mu_0$ are obtained through fitting experimental work function change-measurements for Na on Cu(111) by Fischer et al. [16] in the low-coverage region to Eq. (S5) (see Fig. S6). We obtain $\alpha = 46.6$ Å$^3$ and $\mu_0 = 7.8$ D. This finally results in the effective dipole moment

$$\mu_{\mathrm{eff}}(\Theta) = \frac{\mu_0}{1 + 9\alpha\left(n_0\Theta\right)^{3/2}}. \tag{S6}$$

In our simulations, we truncate the resulting pairwise forces

$$\mathbf{F}^{\mathrm{KL}}_{ij} = -\nabla V^{\mathrm{KL}}_{ij}(\mathbf{r}) = 6\frac{\mu^2_{\mathrm{eff}}(\Theta)}{r^5_{ij}}\mathbf{r}_{ij} \tag{S7}$$

at a cut-off distance of 20 Å.

### C. Langevin Equation and Numerical Propagation

As routinely applied in the context of analyzing $^3$He-SE measurements [17–19], we simulate the actual adsorbate motion using Langevin dynamics that incorporate both the dynamical interaction with substrate phonons as well as with *eh*-pairs equivalently as coupling to an implicit heat bath. Within the Markov approximation the adsorbate dynamics is then determined by

$$m\frac{d^2\mathbf{R}}{dt^2} = -\nabla V_{\mathrm{int}}(\mathbf{R}) - \eta\frac{d\mathbf{R}}{dt} + \mathcal{F}(t). \tag{S8}$$

Here, $\mathbf{R}$ denotes the combined adsorbate coordinates vector, $m$ is the adsorbate mass and $V_{\mathrm{int}}(\mathbf{R})$ is the total interaction potential that includes the Na-Cu(111) interactions $V$ as well as all pairwise repulsive interactions $V^{\mathrm{KL}}$. The fluctuating forces $\mathcal{F}(t)$ are modeled as Gaussian white noise with zero mean and a variance that is related to the friction coefficient $\eta$ and the temperature $T$ by the fluctuation dissipation theorem such that $\langle \mathcal{F}(t)\mathcal{F}(t')\rangle = 2\eta k_{\mathrm{B}} T \delta(t - t')$.

The actual time-propagation of Eq. (S8) is done using the modified velocity Verlet algorithm proposed by Bussi and Parinello [20]. The latter allows to conveniently control the error due to the time discretization of the stochastic Langevin equation by monitoring drifts in the effective energy [20, 21]. With the chosen time step of $\Delta t = 5$ fs we find the latter to be stable to within 0.1 meV per degree of freedom over $2 \times 10^8$ steps.

### D. Electronic Friction

We address the electronically non-adiabatic contributions to the apparent friction coefficient relying on the local density friction approximation (LDFA) [22, 23]. Within the LDFA, atomic electronic friction coefficients $\eta_{eh\text{-pairs},j}$ are individually evaluated from a reference system of a radially symmetric impurity embedded in a free electron gas of a given position-dependent embedding density $\rho_{\text{emb},j}$

$$\eta_{eh-\text{pairs},j}\left(\mathbf{R}_j\right) = \rho_{\text{emb},j}\left(\mathbf{R}_j\right) k_F \sigma_{\text{tr}}, \quad (S9)$$

where $\sigma_{\text{tr}}$ is the transport cross section that is evaluated from the phase shifts of the respective Kohn-Sham orbitals at the Fermi momentum $k_F$ [24–27]. Regarding the density of the host free-electron gas in Eq. (S9) we compare a recently proposed atoms-in-molecules (AIM) embedding scheme [28] based on a Hirshfeld decomposition [29] of the self-consistent system electronic density with the independent-atom approximation (IAA) relying on clean-surface embedding densities [22]. For our analysis we achieve a continuous representation of the electronic friction coefficient $\eta_{eh-\text{pairs},j}(\mathbf{R}_j)$ similar to the adsorbate-substrate interaction potential (see Eq. (S2)). A comparison of the arc-lengths of the actual three dimensional minimum energy paths and their respective two dimensional projections for all relevant single jumps yields deviations of $< 0.5\%$ in all cases. We are thus confident to not neglect any relevant information about the non-adiabatic energy dissipation that comes along with the reduction of the dimensionality of the dynamics.

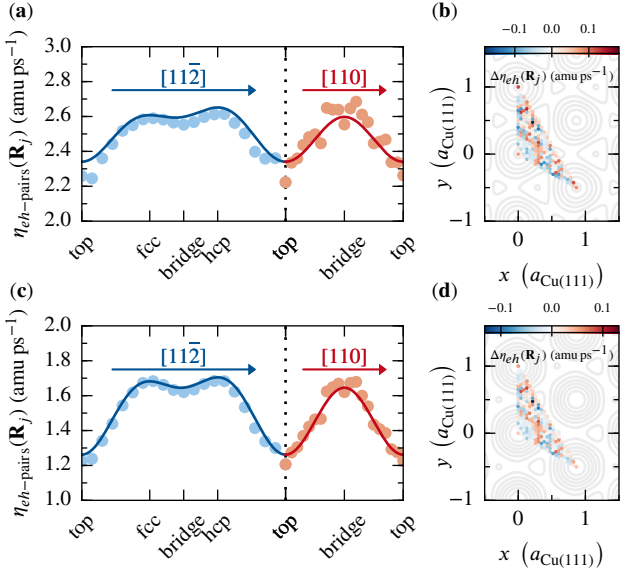

FIG. S7. **(a)** Interpolated Na electronic friction coefficient within the LDFA-AIM model along two high symmetry lines along the Cu(111) surface (drawn lines). Individual input values are shown as filled circles. Note that the majority of deviations is found around the top-site, which is only very rarely visited at the simulated temperatures. **(b)** Deviations $\Delta\eta_{eh}(\mathbf{R}_j) = \eta_{eh\text{-pairs}}(\mathbf{R}_j) - \eta_{eh\text{-pairs}}^{\text{DFT}}(\mathbf{R}_j)$ of the interpolation function from the entire input set of 231 data points in the irreducible wedge of the surface unit cell. **(c)** and **(d)** same as **(a)** and **(b)** but for the LDFA-IAA model.

The respective electronic friction coefficient is highly sensitive to the actual height of the adsorbate due to the exponential decay of the metal electron density above the surface. However, the actual adsorbate height we obtain from our geometry optimizations is determined by minimizing the Hellmann-Feynman forces projected on the adsorbate's z-coordinate. Residual numerical uncertainties that do not affect the latter may still have a small impact on the electronic friction coefficient. This explains the fluctuations of the data points shown in Fig. S7. In order to be less susceptible, we therefore do not only use the electronic friction coefficients at the four high-symmetry sites for the parametrization of the Fourier series, but rather fit a 3 component Fourier expansion (one sine and cosine component, respectively, and a constant contribution) to all 231 data points of our test set from the interaction potential validation using a least-square algorithm. Doing so we obtain a smooth interpolation function with an RMSD value of 0.06 amu ps$^{-1}$ and 0.04 amu ps$^{-1}$ for AIM and IAA, respectively.

### E. Evaluation of the Intermediate Scattering Function

To make the connection between our simulations and the experimental $^3$He-SE measurements we follow a procedure promoted by Ellis and coworkers [11, 18, 30] that avoids the numerically demanding evaluation of the van Hove pair correlation function $G(\mathbf{R}, t)$ for the interacting system [31]. Instead, we rather rely on the kinematic scattering approximation, i.e., we disregard multiple scattering events and consequently evaluate the coherent intermediate amplitudes

$$A_n(\Delta\mathbf{K}, t) = \sum_j^{N_{\text{atoms}}} \exp\left[-i\Delta\mathbf{K} \cdot \mathbf{R}_{n,j}(t)\right] \quad (S10)$$

at each MD time step directly as a superposition of contributions from the trajectories $\mathbf{R}_{n,j}(t)$ generated for an ensemble of $N_{\text{atoms}}$ adatoms in a single run $n$ [32]. The dynamical structure factor

$$S_n(\Delta\mathbf{K}, \omega) = \left|\int_{-\infty}^{\infty} A_n(\Delta\mathbf{K}, t) \exp(-i\omega t)\, dt\right|^2 \quad (S11)$$

is then averaged over several runs and a subsequent inverse Fourier transform finally yields the intermediate scattering function (ISF) [33]

$$I(\Delta\mathbf{K}, t) = \frac{1}{N_{\text{runs}}} \int_{-\infty}^{\infty} \sum_n^{N_{\text{runs}}} S_n(\Delta\mathbf{K}, \omega) \exp(i\omega t)\, d\omega. \quad (S12)$$

Based on this procedure, we subsequently treat both our experimental measurements as well as the simulated (normalized) ISFs on an equal footing by fitting an exponential decay to the latter using the procedure described in Sec. S1 B. Here, special care has to be taken to run trajectories long enough, i.e., well beyond the decay of the ISF to ensure an adequate estimate of the respective decay rate $\alpha(\Delta\mathbf{K})$ that is free of any boundary effects imposed by the numerical Fourier transforms.

### F. Sensitivity Analysis

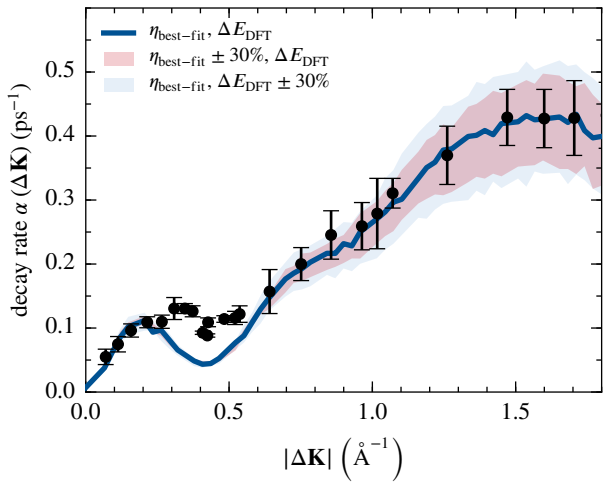

FIG. S8. Experimentally measured decay rates $\alpha(\Delta \mathbf{K})$ along the $[11\bar{2}]$ surface direction as opposed to those extracted from simulations with a best-fit friction coefficient of $\eta = 10$ amu ps$^{-1}$ (solid blue line). The red corridor indicates the range of respective decay rates in simulations with $\eta$ varied by $\pm 30\%$, whereas the blue corridor indicates simulations where the hollow-bridge diffusion barrier as obtained from DFT calculations has been changed by $\pm 30\%$.

Our estimate for the total friction coefficient $\eta$ naturally depends on the quality of the underlying interaction potential. Even though the latter is evaluated on an *ab initio* basis, there are numerical and conceptual approximations inherently included. In order to assess this dependence, we varied the relevant hollow-bridge diffusion barrier in our simulations by up to $\pm 30\%$, which should more than account for values obtained with different sets of pseudo potentials or exchange-correlation functionals. The hollow-top amplitude (see Tab. S1) was left unchanged. Fortunately, given the best-fit friction coefficient, the variations of our so-calculated decay rates (see Fig. S8) are very consistent with the corridor we obtain by varying the friction coefficient as outlined in the main text, and which we already consider as our uncertainty anyway.